\documentclass[useAMS,usenatbib]{mn2e}
\pdfoutput=1
\usepackage{graphicx}

\title[Measures of star formation rates from Infrared ({\it Herschel}) and ultraviolet ({\it GALEX}) emissions of galaxies]{Measures of star formation rates from Infrared ({\it Herschel})  and UV ({\it GALEX}) emissions of galaxies in the HerMES fields}
\author[V.~Buat et al.]
{\parbox{\textwidth}{V.~Buat,$^{1}$\thanks{E-mail: \texttt{veronique.buat@oamp.fr}}
E.~Giovannoli,$^{1}$
D.~Burgarella,$^{1}$
B.~Altieri,$^{2}$
A.~Amblard,$^{3}$
V.~Arumugam,$^{4}$
H.~Aussel,$^{5}$
T.~Babbedge,$^{6}$
A.~Blain,$^{7}$
J.~Bock,$^{7,8}$
A.~Boselli,$^{1}$
N.~Castro-Rodr{\'\i}guez,$^{9,10}$
A.~Cava,$^{9,10}$
P.~Chanial,$^{6}$
D.L.~Clements,$^{6}$
A.~Conley,$^{11}$
L.~Conversi,$^{2}$
A.~Cooray,$^{3,7}$
C.D.~Dowell,$^{7,8}$
E.~Dwek,$^{12}$
S.~Eales,$^{13}$
D.~Elbaz,$^{5}$
M.~Fox,$^{6}$
A.~Franceschini,$^{14}$
W.~Gear,$^{13}$
J.~Glenn,$^{11}$
M.~Griffin,$^{13}$
M.~Halpern,$^{15}$
E.~Hatziminaoglou,$^{16}$
S.~Heinis,$^{1}$
E.~Ibar,$^{17}$
K.~Isaak,$^{13}$
R.J.~Ivison,$^{17,4}$
G.~Lagache,$^{18}$
L.~Levenson,$^{7,8}$
C.J.~Lonsdale,$^{19}$
N.~Lu,$^{7,20}$
S.~Madden,$^{5}$
B.~Maffei,$^{21}$
G.~Magdis,$^{5}$
G.~Mainetti,$^{14}$
L.~Marchetti,$^{14}$
G.E.~Morrison,$^{22,23}$
H.T.~Nguyen,$^{8,7}$
B.~O'Halloran,$^{6}$
S.J.~Oliver,$^{24}$
A.~Omont,$^{25}$
F.N.~Owen,$^{19}$
M.J.~Page,$^{26}$
M.~Pannella,$^{19}$
P.~Panuzzo,$^{5}$
A.~Papageorgiou,$^{13}$
C.P.~Pearson,$^{27,28}$
I.~P{\'e}rez-Fournon,$^{9,10}$
M.~Pohlen,$^{13}$
D.~Rigopoulou,$^{27,29}$
D.~Rizzo,$^{6}$
I.G.~Roseboom,$^{24}$
M.~Rowan-Robinson,$^{6}$
M.~S\'anchez Portal,$^{2}$
B.~Schulz,$^{7,20}$
N.~Seymour,$^{26}$
D.L.~Shupe,$^{7,20}$
A.J.~Smith,$^{24}$
J.A.~Stevens,$^{30}$
V.~Strazzullo,$^{19}$
M.~Symeonidis,$^{26}$
M.~Trichas,$^{6}$
K.E.~Tugwell,$^{26}$
M.~Vaccari,$^{14}$
E.~Valiante,$^{15}$
I.~Valtchanov,$^{2}$
L.~Vigroux,$^{25}$
L.~Wang,$^{24}$
R.~Ward,$^{24}$
G.~Wright,$^{17}$
C.K.~Xu$^{7,20}$ and
M.~Zemcov$^{7,8}$}\vspace{0.4cm}\\
\parbox{\textwidth}{$^{1}$Laboratoire d'Astrophysique de Marseille, OAMP, Universit\'e Aix-marseille, CNRS, 38 rue Fr\'ed\'eric Joliot-Curie, 13388 Marseille cedex 13, France\\
$^{2}$Herschel Science Centre, European Space Astronomy Centre, Villanueva de la Ca\~nada, 28691 Madrid, Spain\\
$^{3}$Dept. of Physics \& Astronomy, University of California, Irvine, CA 92697, USA\\
$^{4}$Institute for Astronomy, University of Edinburgh, Royal Observatory, Blackford Hill, Edinburgh EH9 3HJ, UK\\
$^{5}$Laboratoire AIM-Paris-Saclay, CEA/DSM/Irfu - CNRS - Universit\'e Paris Diderot, CE-Saclay, pt courrier 131, F-91191 Gif-sur-Yvette, France\\
$^{6}$Astrophysics Group, Imperial College London, Blackett Laboratory, Prince Consort Road, London SW7 2AZ, UK\\
$^{7}$California Institute of Technology, 1200 E. California Blvd., Pasadena, CA 91125, USA\\
$^{8}$Jet Propulsion Laboratory, 4800 Oak Grove Drive, Pasadena, CA 91109, USA\\
$^{9}$Instituto de Astrof{\'\i}sica de Canarias (IAC), E-38200 La Laguna, Tenerife, Spain\\
$^{10}$Departamento de Astrof{\'\i}sica, Universidad de La Laguna (ULL), E-38205 La Laguna, Tenerife, Spain\\
$^{11}$Dept. of Astrophysical and Planetary Sciences, CASA 389-UCB, University of Colorado, Boulder, CO 80309, USA\\
$^{12}$Observational  Cosmology Lab, Code 665, NASA Goddard Space Flight  Center, Greenbelt, MD 20771, USA\\
$^{13}$Cardiff School of Physics and Astronomy, Cardiff University, Queens Buildings, The Parade, Cardiff CF24 3AA, UK\\
$^{14}$Dipartimento di Astronomia, Universit\`{a} di Padova, vicolo Osservatorio, 3, 35122 Padova, Italy\\
$^{15}$Department of Physics \& Astronomy, University of British Columbia, 6224 Agricultural Road, Vancouver, BC V6T~1Z1, Canada\\
$^{16}$ESO, Karl-Schwarzschild-Str. 2, 85748 Garching bei M\"unchen, Germany\\
$^{17}$UK Astronomy Technology Centre, Royal Observatory, Blackford Hill, Edinburgh EH9 3HJ, UK\\
$^{18}$Institut d'Astrophysique Spatiale (IAS), b\^atiment 121, Universit\'e Paris-Sud 11 and CNRS (UMR 8617), 91405 Orsay, France\\
$^{19}$National Radio Astronomy Observatory, P.O. Box O, Socorro NM 87801, USA\\
$^{20}$Infrared Processing and Analysis Center, MS 100-22, California Institute of Technology, JPL, Pasadena, CA 91125, USA\\
$^{21}$School of Physics and Astronomy, The University of Manchester, Alan Turing Building, Oxford Road, Manchester M13 9PL, UK\\
$^{22}$Institute for Astronomy, University of Hawaii, Honolulu, HI 96822, USA\\
$^{23}$Canada-France-Hawaii Telescope, Kamuela, HI, 96743, USA\\
$^{24}$Astronomy Centre, Dept. of Physics \& Astronomy, University of Sussex, Brighton BN1 9QH, UK\\
$^{25}$Institut d'Astrophysique de Paris, UMR 7095, CNRS, UPMC Univ. Paris 06, 98bis boulevard Arago, F-75014 Paris, France\\
$^{26}$Mullard Space Science Laboratory, University College London, Holmbury St. Mary, Dorking, Surrey RH5 6NT, UK\\
$^{27}$Space Science \& Technology Department, Rutherford Appleton Laboratory, Chilton, Didcot, Oxfordshire OX11 0QX, UK\\
$^{28}$Institute for Space Imaging Science, University of Lethbridge, Lethbridge, Alberta, T1K 3M4, Canada\\
$^{29}$Astrophysics, Oxford University, Keble Road, Oxford OX1 3RH, UK\\
$^{30}$Centre for Astrophysics Research, University of Hertfordshire, College Lane, Hatfield, Hertfordshire AL10 9AB, UK}}

\begin{document}

\date{}

\pagerange{\pageref{firstpage}--\pageref{lastpage}} \pubyear{2002}

\maketitle

\label{firstpage}
\clearpage
\begin{abstract}
 {The reliability of infrared (IR) and ultraviolet (UV)  emissions to measure star formation rates in galaxies is investigated for a large sample of galaxies observed with the SPIRE and PACS instruments  on  {\it Herschel}  as part of  the HerMES  project. We build flux-limited 250 $\mu$m samples of sources at redshift $z<1$,  cross-matched with the {\it {\it Spitzer}}/MIPS and {\it GALEX} catalogues.  About  60 $\%$ of the Herschel sources are detected in UV.
 The  total  IR luminosities, $L_{\rm IR}$, of the sources are estimated using a SED-fitting code that fits to fluxes between 24 and 500 $\mu$m.  Dust attenuation is discussed on the basis of commonly-used diagnostics:  the $ L_{\rm IR}/L_{\rm UV}$ ratio  and the slope, $\beta$, of the UV continuum. A mean dust attenuation $A_{\rm UV}$ of $\simeq 3$ mag is  measured in the samples.  $ L_{\rm IR}/L_{\rm UV}$   is found to correlate with $L_{\rm IR}$. Galaxies with $ L_{\rm IR} > 10 ^{11} \rm L_{\odot}$  and $0.5< z<1$ exhibit a mean dust attenuation $A_{\rm UV}$ about $ 0.7$ mag lower than that found for their local counterparts, although with a large dispersion.  Our galaxy samples span  a large range of $\beta$ and $L_{\rm IR}/L_{\rm UV}$ values which, for the most part, are distributed between the ranges  defined by the relations found locally for starburst and normal star-forming galaxies. As a consequence the  recipe commonly applied to  local starbursts is found to overestimate the dust attenuation correction in our galaxy sample by a factor $\sim 2-3$. 
The SFRs deduced from  $L_{\rm IR}$ are found to account for   about $90\%$  of the total SFR; this percentage drops to 71$\%$   for  galaxies with $\rm SFR < 1 M_{\odot} yr^{-1}$ (or $L_{\rm IR} < 10^{10} \rm  L_{\odot}$). For these faint objects, one needs to combine UV and IR emissions to obtain an accurate measure of the SFR. }

\end{abstract}


\section{Introduction}
Far-infrared (IR) and ultraviolet (UV) luminosities are commonly used to estimate the current star formation rate (SFR) in galaxies since both emissions are expected to come from young stars. 
By combining observed IR and UV luminosities one can therefore make an energetic budget and derive  an accurate measure of the total SFR in star-forming galaxies \citep[e.g.][]{iglesias04,elbaz07}. It is often the case that only UV or IR data exist, however, and so the question remains  of how reliably  one can  determine SFRs from UV or IR alone. \\
SFRs derived from dust emission are based on estimates of the total IR luminosity ($L_{\rm IR}=L[8-1000{\rm \mu m}]$).  This measure is accurate in the nearby Universe thanks to the observations of {\it IRAS} and {\it {\it Spitzer}}, both of which sampled the wavelength range close to the peak emission of the dust. {\it Spitzer} was also used to observe galaxies up to $z\approx2$ in the mid-IR, however extrapolation to the total IR luminosity remains uncertain and relies on relations determined from nearby galaxy populations. By observing in the rest-frame far-infrared of galaxies, {\it Herschel} \citep{pilbratt10} \footnote{Herschel is an ESA space observatory with science instruments provided by Principal Investigator consortia. It is open for proposals for observing time from the worldwide astronomical community} allows us to measure accurately this IR bolometric luminosity over a continuous and wide range of redshift.\\
The main issue  when using  UV emission to estimate SFRs is the effect of  dust attenuation. The $L_{\rm IR}/L_{\rm UV}$ ratio has been identified as a very powerful estimator of dust attenuation in star forming galaxies \citep[e.g.][]{gordon00,buat05}.
 Dust attenuation diagnostics based on UV data alone must be used when UV and IR rest-frame data are not simultaneously available.   \citet{Meurer99} found a relation between the slope of the rest-frame UV continuum $\beta$ (defined as $ f_\lambda ({\rm erg ~cm^{-2} s^{-1} nm^{-1}}) \propto \lambda^\beta$  for $\lambda > 120 $ nm) and  dust attenuation traced by $L_{\rm IR}/L_{\rm UV}$ for local starburst galaxies observed by {\it IUE} and {\it IRAS}.  This  local starburst relation is also widely used to estimate dust attenuation in UV-selected galaxies at high redshift \citep{burgarella07, reddy08}.  In the local Universe  luminous IR  galaxies  (LIRGs with $L_{\rm IR} > 10^{11} \rm L\odot$) are found to  roughly follow the local starburst law \citep{takeuchi10, howell10}; normal  forming galaxies, less active in star formation than starburst ones, do not   follow this relation, and the spectral slope $\beta$ is found to depend on a number of parameters that are not solely related to dust attenuation \citep[e.g.][]{buat05,boissier07, kong04}. 
Checking the validity of $\beta$ as a tracer of  dust attenuation on large samples of galaxies at different redshifts is therefore a key to both understanding dust attenuation processes in galaxies, and to being able to correct accurately for extinction.\\
In this paper we will analyse the IR and UV properties of galaxies over the redshift range, $0<z<1$. This is the first time accurate estimates of the total dust emission have  been determined over  a continuous range of redshift by using several IR measurements  including  new {\it Herschel} data and which can be compared to the observed UV emission.    

\section{Data}
 Observations of the Lockman field were made with {\it Herschel} as part of the {\it Herschel}  Multi-Tiered Extragalactic Survey (HerMES-\citet{oliver10}) \footnote { hermes.sussex.ac.uk}.  This field was also observed by 
{\it GALEX} \citep{martin05}  in FUV ($\lambda \simeq$153 nm) and NUV($\lambda \simeq$ 231 nm) as part of the Deep Imaging Survey (DIS). Only sources with $z<1$  are detected by {\it GALEX} due to absorption by the Lyman Break.\\
 Our first parent sample is based on the shallow observations of the Lockman SWIRE field (10h 45m 00s +58d 00m 00s,  $218'\times218'$ ) with SPIRE \citep{griffin10}. 
 We use the HerMES cross-identified  catalogue  of \citet{roseboom10} which is based on a linear inversion method  using  the positions of sources detected in the {\it {\it Spitzer}} 24 $\mu$m surveys. We select    7435 sources. 
that are detected at 5$\sigma$ or above at 250 $\mu$m, corresponding to a limit of 23 mJy \citep{roseboom10}. A redshift has been  assigned to  3862 sources. Where available, we use spectroscopic redshifts (taken from the NASA/IPAC Extragalactic Database), otherwise photometric redshifts from  the SDSS/DR7 \citep{abazajian09}  and the SWIRE survey   \citep{Rowan-Robinson08} are used by order of priority. We identify 3824 galaxies with $z<1$, of which  30$\%$  have spectroscopic redshifts.  This sample is then  cross-matched with the {\it GALEX} detections in the NUV with a tolerance radius of 2 arcsec (based on {\it {\it Spitzer}} /IRAC  coordinates): 2426 galaxies  are detected in NUV with  $z<1$.   The UV slope $\beta$ can be calculated from the $FUV-NUV$ colour for sources with $z<0.3$. Of the 1542 objects at  $z<0.3$, 1250 are detected in the NUV and 943 have both FUV and NUV fluxes that are of sufficient accuracy ($\Delta m <0.1$mag) to  enable  reliable estimates  of  $\beta$. \\
Although $\sim 40\%$ of the Lockman SWIRE sources selected at 250 $\mu$m are not detected in the NUV it is quite difficult to study the non detections since not less than 21 GALEX fields with different depths are needed to cover the HerMES field. Therefore we also consider the {\it Herschel} deep observations of the Lockman North field ( 10h46m00s +59d01m00s,  $35'\times 35'$, 
5$\sigma$ limit at 250 $\mu$m corresponding to 8 mJy \citep{roseboom10}): 70$\%$ of this  field is  covered by only 3 {\it GALEX} fields and a check of each individual source by hand  is therefore possible. A second advantage of the Lockman-North field is that it was also observed with PACS \citep{poglitch10} and data at 110 and 160 $\mu$m are available.
We again use the cross-matched catalogue  of \citet{roseboom10} and perform a selection and a {\it GALEX} cross-match similar to that made for the Lockman SWIRE field. Photometric redshifts were  determined by \citet{strazzullo10} from visible, NIR and IRAC data. 246 sources   are selected at 250 $\mu$m, of which 129 galaxies are detected in the  NUV. The 58 sources that are considered as non-detections in the  NUV are assigned a limiting magnitude of NUV= 24.4  \citep{morissey05}. We consider  as  UV upper limits, objects with no  GALEX source  within 6 arcsec of  the SPIRE position  (for distances less than 6 arcsec the full NUV PSF falls into the 250$\mu$m one)).\\
Total  IR luminosities,  $ L_{\rm IR}$, are  calculated using  the CIGALE code \citep{noll09}    with  \citet{dale02}  and  \citet{Siebenmorgen07} IR templates. A Bayesian analysis is performed to deduce physical parameters from the fitting of spectral energy distributions of galaxies. 
We restrict the analysis to the thermal IR ($\rm \ge 24 \mu m$ data), and defer an analysis of the full UV-to-FIR SEDs to a later paper.   For each source we consider all the available data between 24 and 500 $\mu$m: by construction we will always have at least two data points.
We keep only objects for which  the reduced chi-squared of the fit between the  best template and the dataset is lower than 10  (98$\%$ of the initial samples).  Typical errors on estimated $L_{\rm IR}$ are found lower than 0.1 dex. UV luminosities $L_{\rm UV}$ are derived at 153 nm rest-frame (FUV) by interpolating FUV and NUV fluxes, a mean $FUV-NUV$ colour is used when FUV data are not available, and are defined  as the quantities $\nu L_{\nu}$. Both $L_{\rm IR}$ and $L_{\rm UV}$ are expressed  in solar units. All magnitudes are given in the AB system. We assume that $\Omega_m = 0.3$, $\Omega_{\Lambda} = 0.7$, and $ H_0 = 70 {\rm~ km~ s^{-1}~ Mpc^{-1}}$.  

\section{Dust attenuation as traced by $L_{\rm IR}/L_{\rm UV}$}
 \begin{figure}
\includegraphics[width=85mm]{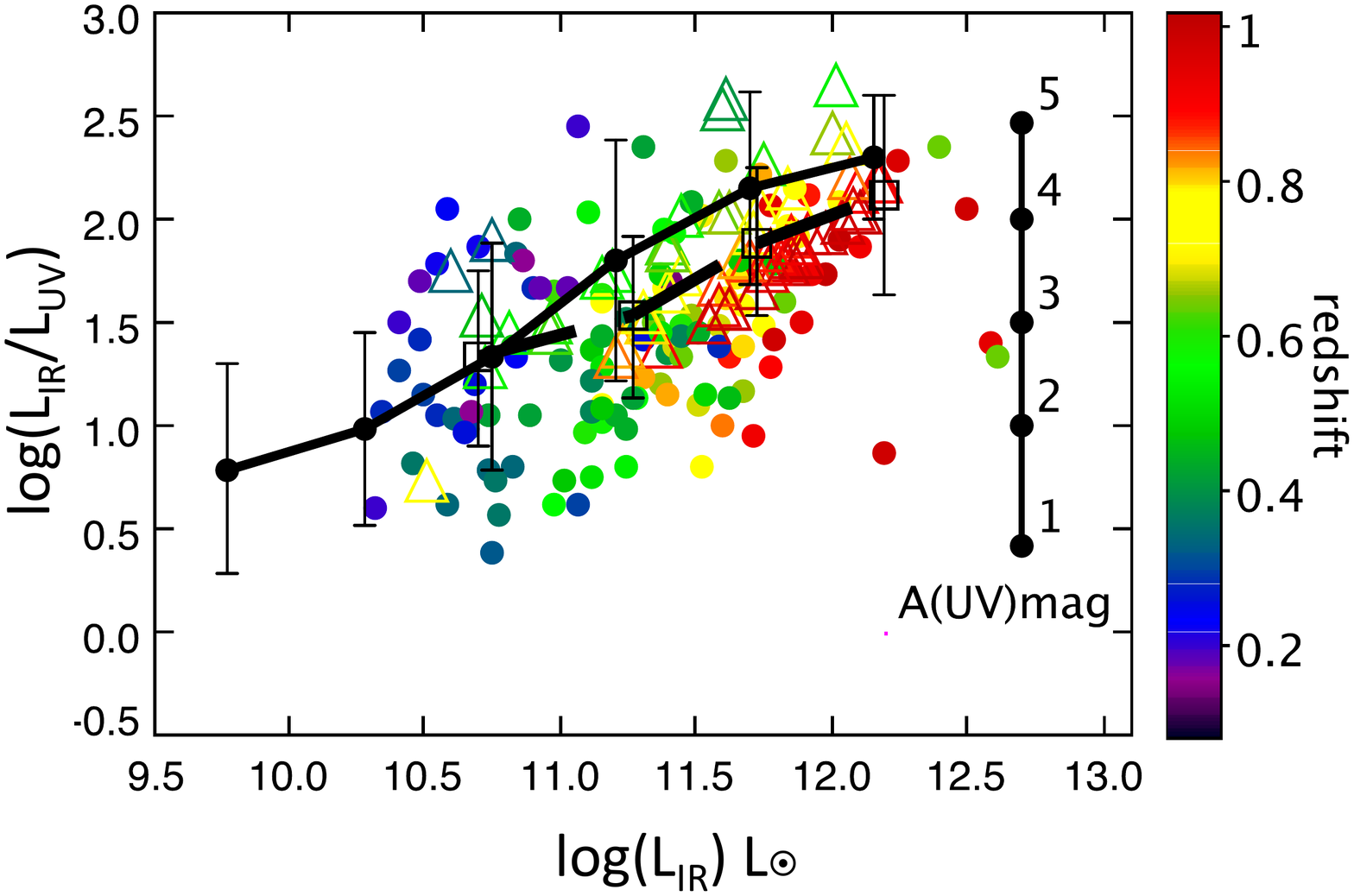}
   \caption{$L_{\rm IR}/L_{\rm UV}$ vs. $L_{\rm IR}$ for sources detected in the Lockman North in the redshift range $0<z<1$ (the redshift is colour-coded).  Lower limits of $L_{\rm IR}/L_{\rm UV}$ are plotted as triangles for galaxies not detected in UV. The filled circles  and  solid line  correspond to the relation found for local IR selected galaxies \citep{buat07a}, the empty squares  and dashed line to the mean values per bin of IR luminosity obtained in  the present study.  Errors bars correspond to the scatter of the data (1 rms). The A(UV) scale is obtained using the calibration of \citet{buat05}. The luminosity bins and mean redshift in the bins are: $\log(L_{\rm IR}) < 11$ and $<z>=0.4$, $11\le \log(L_{\rm IR}) < 11.5$ and $<z>=0.55$, $11.5 \le \log(L_{\rm IR}) < 12$ and $<z>=0.8$, $\log(L_{\rm IR}) \ge 12 $ and $<z>=0.9$.}
              \label{IRX-Ldust}%
    \end{figure}
 The ratio of $L_{\rm IR}/L_{\rm UV}$ is a powerful measure of dust attenuation and is known to correlate with the bolometric luminosity of galaxies in the nearby universe \citep[e.g.][]{buat07a}. {\it {\it Spitzer}}  observations extended the analysis out  to $z\simeq 0.7$ \citep{zheng07,buat07b,xu07}.  
 Here we re-investigate this relation using the Lockman North sample in the redshift range $0<z<1$. We choose this field  in order to be able to discuss lower limits for galaxies not detected in UV. Shown in Fig.~\ref{IRX-Ldust} is a plot of  $ L_{\rm IR}/L_{\rm UV}$ as a function of $L_{\rm IR}$.  The mean dust attenuation is $A(UV) = 3\pm 1$ mag for the whole sample (using the calibration of  \citet{buat05}). As expected, there is a  general increase of $L_{\rm IR}/L_{\rm UV}$ with $L_{\rm IR}$.  We measure the mean  $L_{\rm IR}/L_{\rm UV}$ ratio within several bins of $\log(L_{\rm IR})$ (using the Kaplan-Meier estimator  to account for lower limits). The locus of galaxies at low z appears to be consistent with the $z=0$ relations  \citep{buat07a}, however at $z> 0.5$   $L_{\rm IR}/L_{\rm UV}$   is found to be lower by $\sim 0.3$ dex. This difference at higher $z$ is of low statistical significance  (of the order of the 1 $\sigma$ error bars) and could be affected by the lower-limit corrections but if it is real this implies a decrease in dust attenuation by $\sim 0.7$ mag in the FUV for distant, luminous IR-selected galaxies which is consistent with dust attenuation found for LIRGs at $z=0.7$ by \citet{buat07b}. 
 A decrease in dust attenuation is possibly linked to  a lower metallicity in high-redshift systems. Indeed    \citet{reddy10} found a correlation between oxygen abundance and dust attenuation in $z\sim 2$ star forming galaxies. The morphological evolution of LIRGs from mergers to more normal galaxies as $z$ increases can also imply a decrease of dust attenuation  \citep[e.g.] [and references therein]{buat07b}. Because of  a larger gas mass fraction in distant galaxies, the star formation could be spread out over larger,  more diffuse and less extinguished  regions than in local LIRGs.  A decrease of dust attenuation was also reported by \citet{daddi07} for Ultraluminous IR galaxies  between $z=0$ and  $z \sim 2$ associated to a longer lived star formation for the higher redshift sources.
     \begin{figure}
   \centering
 \includegraphics[width=90mm]{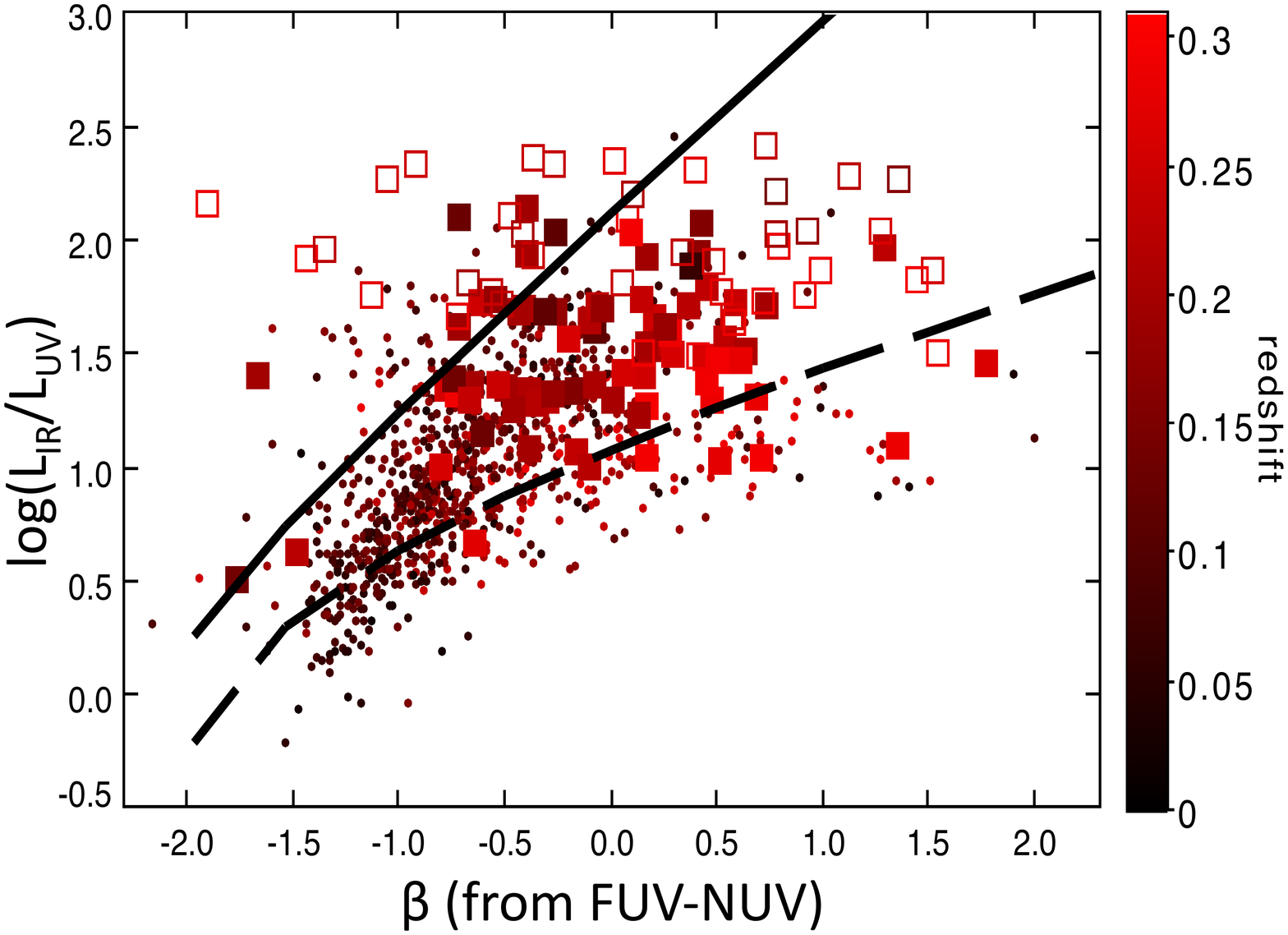}
   \caption{$L_{\rm IR}/L_{\rm UV}$ vs. $\beta$ (the slope of the UV continuum) for galaxies in the Lockman SWIRE field that are also detected in FUV and NUV ($z<0.3$): dots represent galaxies with $\rm L_{\rm IR}<10^{11} L_{\odot}$ and a robust measure of $\beta$;  filled squares represent LIRGs ( $\rm L_{\rm IR}>10^{11} L_{\odot}$)  with a robust measure of $\beta$, and LIRGs with less reliable $\beta$ values are represented by empty squares. The redshift of sources is colour-coded. The superposed lines trace the relation found for local starburst galaxies (solid  line) and  for local normal star-forming galaxies (dashed  line).}
              \label{IRX-beta}%
    \end{figure}
\section{Dust attenuation as traced by the UV slope}
A common way to estimate dust attenuation in the UV in the absence of IR data is to use the slope $\beta$ of the UV continuum. 
To estimate $\beta$ from the $FUV-NUV$ colour we focus on galaxies with $z<0.3$ in the Lockman SWIRE sample and apply the  \citet{kong04} recipe. 
A measurement error of 0.1 mag in FUV and NUV  translates into an error of 0.3 for  $\beta$. Therefore  we restrict the analysis (otherwise specified) to sources measured in FUV and NUV with an accuracy better than 0.1 mag.
We have checked that considering the whole sample of  galaxies detected in FUV and NUV does not change the results, only adding a larger dispersion in the diagrams.  
Shown in Fig.~\ref{IRX-beta}  is a plot of $L_{\rm IR}/L_{\rm UV}$ as a function of $\beta$, together with local relations for starbursts  \citep{kong04} and optically selected  star-forming galaxies  \citep{boissier07}  are  superposed.  Galaxies from our IR-selected sample exhibit  a wide range of $\beta$ and $L_{\rm IR}/L_{\rm UV}$ values as already found  in the nearby universe \citep{buat05, seibert05, takeuchi10}. The majority of  the galaxies lie  between the two relations  i.e. have a dust attenuation (as defined by $L_{\rm IR}/L_{\rm UV}$)  for a given $\beta$ that is larger than that found in local normal star-forming galaxies but lower than in local starbursts.   From the 164 LIRGs of our initial sample of 1542 sources  at $z<0.3$ in the Lockman SWIRE field,  125  are detected in NUV and FUV,  of which  78 have a reliable estimate of $\beta$ ($\Delta m <0.1$mag in FUV and NUV) and are shown in Fig.~\ref{IRX-beta}. 67 of these 78 LIRGs  (i.e. $87\%$)    are   found to lie below the local  starburst relation. The fraction drops to 77$\%$ for the 125 LIRGs detected in both FUV and NUV without any restriction on the measurement accuracy.  11\% of the galaxies (including 10 LIRGs)  detected in NUV  have no FUV  detection (for these we adopt a limiting FUV magnitude of 24.8   \citep{morissey05} ; if we  consider only sources with NUV magnitudes with $\Delta NUV < 0.1$, all these sources are found below the starburst law. This estimate might be uncertain given the large number of {\it GALEX} fields with different exposure times. As quoted in Section 2, 23$\%$ of the initial sample at $z < 0.3$th is detected neither in NUV nor FUV: the limiting magnitude of $NUV=24.4$  roughly  corresponds to  a limit on  $L_{\rm IR}/L_{\rm UV}$ of $\simeq 200$  in the redshift range $0<z<0.3$ (for the lowest IR luminosities at each redshift). Therefore we miss at most 23$\%$ of galaxies with $\log(L_{\rm IR}/L_{\rm UV})>2.3$ (note that fewer than 23$\%$ may be genuine non-detections as discussed for the Lockman North field). A detailed discussion of  UV non detections  is deferred to a future paper. Nevertheless, even if we account for galaxies not detected in the NUV and FUV,  a substantial fraction of all the IR-selected galaxies is found below the local  starburst law. Dust attenuation in these galaxies  is {\it overestimated} if the recipe for starburst galaxies based on the UV spectral slope is applied.\\
The most distant galaxies of our sample, including LIRGs, are found to depart strongly from the starburst law (see Fig.~\ref{IRX-beta}). 
$\beta$ is calculated from the observed $FUV-NUV$ colour: the wavelengths sampled go from 150 and 230 nm at z=0 to 118 and 180 nm at z=0.3. 
Therefore, the power law model for the UV continuum assumed to calculate $\beta$ might not be valid  for the whole 118-230 nm range and  for  dust  attenuations  ranging from 0.5 to 5 mag in UV. The uncertainty  can also be amplified because of the rather large bandwidths of the GALEX filters \citep{morissey05}. Variations in dust  properties and star versus dust geometry are known to induce large dispersions in the $L_{\rm IR}/L_{\rm UV}-\beta$  diagram \citep{gordon00,  boquien09}). Our selection at 250 $\mu$m may also lead to more quiescent galaxies than local LIRGs in terms of star formation which is  known to increase $\beta$ for a given $L_{\rm IR}/L_{\rm UV}$ \citep{kong04}.
 At  z=2 \citet{reddy10} found that Lyman Break Galaxies also detected at 24 $\mu$m and with bolometric luminosity lower than $\rm 10^{12}  L_{\odot}$ follow the local starburst law, although with a large dispersion. These galaxies have $\beta < -0.5$ and therefore  are much bluer than the bulk of our sample.  At high redshift the dynamical range in stellar population is narrower and the contribution to the UV continuum of stars not related to the current star formation is likely to be negligible in star forming galaxies at $z>1$.  The spatial distribution of dust and young stars  may also be similar to that found in strong  local starbursts making the  law proposed by \citet{Meurer99} more appropriate for these high redshift objects than for our present sample.

This topic  will be re-investigated in a future work  by fitting the whole SED from the UV to the IR both at low and high redshift. \\

\section{Calibrating measures of Star Formation Rate}

\begin{figure}
   \centering
\includegraphics[width=90mm]{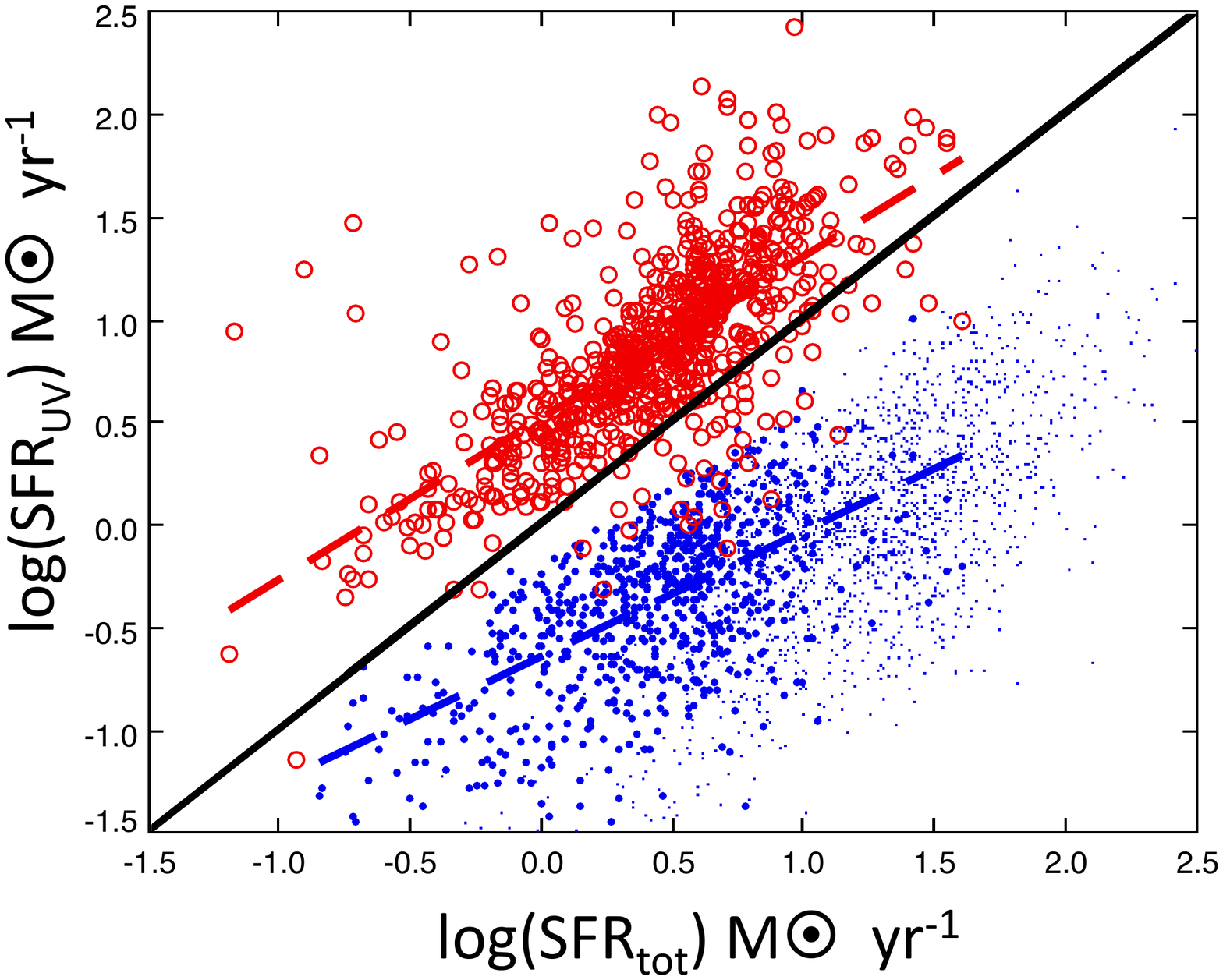}
   \caption{A plot of ${\rm SFR}_{\rm UV}$ vs. ${\rm SFR_{tot}}$, where $ {\rm SFR_{tot}} = {\rm SFR}_{\rm IR}+{\rm SFR}_{\rm UV}$. ${\rm SFR}$s deduced from uncorrected  UV fluxes are plotted with blue  dots, the largest ones correspond to galaxies at $z<0.2$. SFRs deduced from UV fluxes corrected for dust attenuation  with the relation of \citet{Meurer99} are plotted with open green circles for galaxies with $z<0.2$. The linear regression lines between  ${\rm SFR_{tot}}$ and the two estimates of  ${\rm SFR}_{\rm UV}$ are plotted as dashed lines. The black solid line corresponds to equal values on both axes. }
             \label{SFR-UV}%
    \end{figure}

An important check  is to test whether IR emission alone provides a robust estimate of the total ${\rm SFR}$ over the whole range of IR luminosities or if ignoring the direct UV light leads to systematic errors. We  again use the sample derived from the Lockman-SWIRE  field because of the large number of galaxies in the sample.  \\
 We take $ {\rm SFR}_{\rm IR}+{\rm SFR}_{\rm UV}={\rm SFR_{tot}}$ as a reference for the SFR measure 
 where  $\log({\rm SFR}_{\rm IR})_{\rm M_{\odot} yr^{-1}} = \log(L_{\rm IR})_{\rm L_{\odot}}-9.97$ and $\log({\rm SFR}_{\rm UV})_{\rm M_{\odot} yr^{-1}} = \log(L_{\rm UV})_{\rm L_{\odot}}-9.69$. The UV emission is not corrected for dust attenuation and we use the ${\rm SFR}$ calibrations  from \citet{buat08} with the assumption of a constant ${\rm SFR}$ over $10^8$ years and a Kroupa initial mass function \citep{kroupa01}.\\
 \begin{figure}
   \centering
\includegraphics[width=90mm]{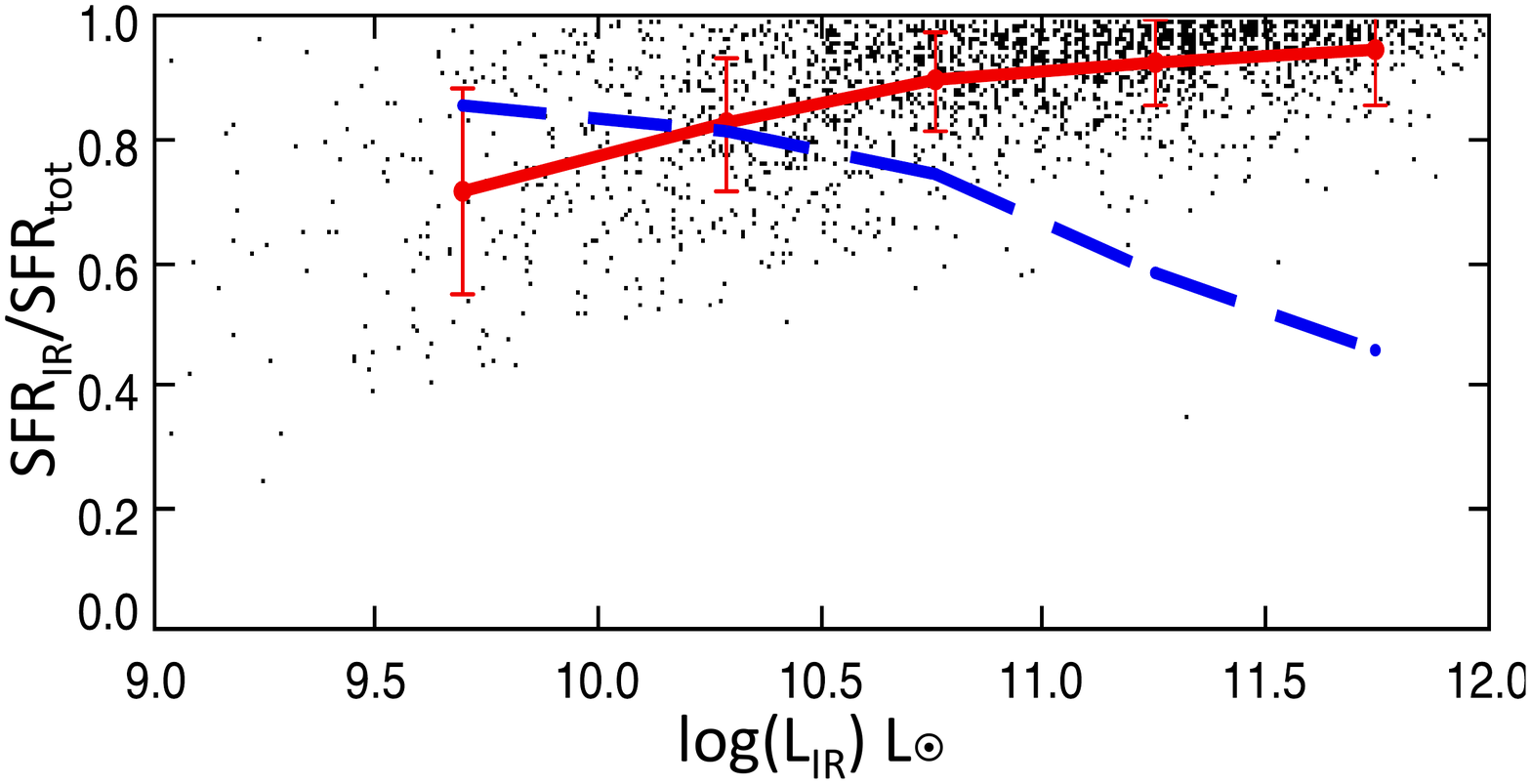}
  \caption{Fraction of SFR measured in IR, ${\rm SFR}_{\rm IR}$  as a function of the IR luminosity, $L_{\rm IR}$. The mean and standard deviation of the values in 5 bins of $\log(L_{\rm IR}$: $<10$, 10-10.5, 10.5-11, 11-11.5, $>11.5$ are shown in red, joined by a solid line to guide the eye.The blue dashed line represents the fraction of galaxies detected in the UV at each value of $\log(L_{\rm IR})$}
\label{SFRfraction}%
    \end{figure}
  Plotted in  Fig.~\ref{SFRfraction} is the ${\rm SFR}$ fraction measured by the  IR luminosity, ${\rm SFR}_{\rm IR}/{\rm SFR_{tot}}$,  as a function of $L_{\rm IR}$, with an additional curve showing the fraction of the galaxies in the sample with that IR luminosity that are detected in the UV. When  the sample is considered as a whole, we find that ${\rm SFR}_{\rm IR}$ measures $\sim 90\%$ of the total SFR. This fraction varies with $L_{\rm IR}$: from 94$ \pm 10\%$ for the most luminous galaxies of our sample ($L_{\rm IR} > 10^{11.5} ~ \rm{ L_{\odot}}$) with  45$\%$ detected in UV, down to  $71\pm 17\%$   when  $L_{\rm IR} < 10^{10} ~\rm{ L_{\odot}}$ or equivalently  ${\rm SFR_{tot}}< 1 \rm {M_{\odot} yr^{-1}}$ and a UV detection rate reaching 85$\%$. This demonstrates that the combination of  obscured and unobscured SFRs  is required to determine accurate SFRs in galaxies with low star formation activity.  Intrinsically faint galaxies detected in rest-frame UV surveys are found  to be a significant component of  luminosity functions  and of the total star formation density at both low and high redshifts  \citep{buat07a,reddy09}. As a consequence, both UV and IR selected samples must be built and their contribution added to measure the total star formation density at a given redshift \citep{iglesias06,reddy09}. \\
Shown in Fig~\ref{SFR-UV} is a comparison of the ${\rm SFR}$ measured in the UV with ${\rm SFR_{tot}}$,  for both uncorrected and corrected  UV luminosities, where the correction used is that of  \citet{Meurer99}: $A_{\rm UV} = 4.43+1.99 \beta$ deduced from their   $L_{\rm IR}/L_{\rm UV}$-$\beta$ relation.  For this correction we only consider  galaxies with $z<0.2$  to narrow the wavelength range used  to calculate    $\beta$. Without dust attenuation corrections, ${\rm SFR}_{\rm UV}$ underestimates the total SFR  by a factor  $\sim 6$ for ${\rm SFR_{tot}}\simeq  1 ~\rm{M_{\odot} yr^{-1}}$ and $\sim 30$  for  ${\rm SFR}_{tot}\simeq \rm 100 ~M_{\odot} yr^{-1}$.  As expected from Fig~\ref{IRX-beta}, with dust corrections based on $\beta$,  SFRs are  overestimated  by a factor  $\sim 2-3$. At  $z \sim1$    \citet{elbaz07}  only obtain ${\rm SFR}  \leq 10 ~\rm M_{\odot} yr^{-1} $   when correcting UV luminosities for dust attenuation based on values of $\beta$. 
They  use the $U-B$  colour to measure $\beta$ which corresponds to the wavelength range 180-225 nm  at $z\sim1$;  which is different from the ranges used in this work and by \citet{Meurer99}.   \citet{burgarella07} and \citet{reddy09}  corrected UV measurements  using the $L_{\rm IR}/L_{\rm UV}$-$\beta$ relation for local starbursts  and found  ${\rm SFR}$ deduced from UV-corrected luminosities  to be in agreement with ${\rm SFR_{tot}}$ out to ~few 100s M$\rm _{\odot} ~yr^{-1}$, allbeit with significant dispersion.   As discussed in section 4, this agreement is likely to be due to the different nature of UV selected galaxies at $z>1$ as compared to the IR selected galaxies analysed in the present  work.

\section*{Acknowledgements}
SPIRE has been developed by a consortium of institutes led by Cardiff University (UK) and including Univ. Lethbridge (Canada); NAOC (China); CEA, OAMP (France); IFSI, Univ. Padua (Italy); IAC (Spain); Stockholm Observatory (Sweden); Imperial College London, RAL, UCL-MSSL, UKATC, Univ. Sussex (UK); and Caltech/JPL, IPAC, Univ. Colorado (USA). This development has been supported by national funding agencies: CSA (Canada); NAOC (China); CEA, CNES, CNRS (France); ASI (Italy); MCINN (Spain); SNSB (Sweden); STFC (UK); and NASA (USA). The data presented in this paper will be released through the {\it Herschel} Database in Marseille HeDaM (http://hedam.oamp.fr/HerMES) This work makes use of TOPCAT http://www.star.bristol.ac.uk/~mbt/topcat/.

\bsp

\label{lastpage}

\end{document}